\documentclass[aps,prl,groupedaddress,twocolumn]{revtex4}

\usepackage{graphicx}
\usepackage{amsmath}
\usepackage{amssymb}
\usepackage{bm}

\newcommand{\ket}[1]{\ensuremath{\left|{#1}\right\rangle}}

\newcommand{\brm}[1]{\ensuremath{\mathbf{#1}}}
\begin{document}


\title{Four-dimensional Quantum Key Distribution using Position-Momentum and Polarization Correlations}
\author{M. P. Almeida}
\affiliation{Instituto de F\'{\i}sica, Universidade Federal do Rio de
Janeiro, Caixa Postal 68528, Rio de Janeiro, RJ 21941-972, Brazil}
\author{S. P. Walborn}
\email[]{swalborn@if.ufrj.br}
\affiliation{Instituto de F\'{\i}sica, Universidade Federal do Rio de
Janeiro, Caixa Postal 68528, Rio de Janeiro, RJ 21941-972, Brazil}
\author{P. H. Souto Ribeiro}

\affiliation{Instituto de F\'{\i}sica, Universidade Federal do Rio de
Janeiro, Caixa Postal 68528, Rio de Janeiro, RJ 21941-972, Brazil}

\date{\today}

\begin{abstract}
We demonstrate experimentally that it is possible to prepare and detect photon pairs created by spontaneous parametric down-conversion which exhibit simultaneous position-momentum and polarization correlations that are adequate to implement a four-dimensional key distribution protocol. 
\end{abstract}


\maketitle
Quantum key distribution (QKD) is currently one of the most promising areas of quantum 
information.  To date there are many QKD protocols \cite{gisin02}, beginning with the 
seminal work of Bennett and Brassard \cite{bb84}, commonly known as the BB84 protocol.  
While BB84 uses single particles, also well-known is the Ekert protocol \cite{ekert91}, 
which utilizes entangled pairs to distribute a random and secure key.   
Bennett, Brassard and Mermin (BBM) have adapted the BB84 protocol to entangled particles, 
and shown that the BB84 and Ekert protocols are essentially equivalent \cite{bennett92b}. 
\par
The object of QKD is to establish a random and secret keystring between two separate
parties $A$ (Alice) and $B$ (Bob). 
Two important goals in experimental and theoretical research in QKD are (\textit{i}) 
increasing the security and (\textit{ii}) increasing the bit transmission rate between 
$A$ and $B$.  It has been shown that they can achieve both of these goals by using
higher-dimensional qudits (with $d$ orthogonal states) \cite{bechmann00a,bourennane01}.
For example, for the $d=4$ case (qu-quarts), using a generalized BB84 or BBM protocol, the
transmission rate would be improved from $\frac{1}{2}$ (for qubits) to $1$ bit per photon (BB84)
or photon pair (BBM). The error rate due to eavesdropping based on the intercept-resend strategy would also increase from $\frac{1}{4}$ to $\frac{3}{8}$, therefore
improving the security.
\par
  Photon pairs obtained from parametric down-conversion have played a crucial role in the experimental investigation of the fundamentals and applications of quantum mechanics \cite{bouwmeester00}.  On the one hand, most experiments in quantum information and computation use photons pairs entangled in the polarization degree of freedom (DOF) \cite{kwiat95,kwiat99}.  On the other hand, position-momentum entanglement has been investigated
and utilized for more than ten years in the context of quantum imaging \cite{ribeiro95,monken98,fonseca99b} and more recently, it was shown that down-converted photon pairs are indeed entangled in position-momentum \cite{howell04,dangelo04} through the violation of a separability criterion \cite{duan00,mancini02}.    Photons are the obvious choice for QKD, due to the ease in which approximate single photon pulses or entangled photon pairs can be produced, manipulated and 
transmitted. Using entangled photons, QKD has been demonstrated using 
polarization \cite{jennewein00,naik00}, time-bin \cite{tittel00} and transverse 
position-momentum \cite{almeida05}.
\par
In this letter, we show that it is possible to create and measure 
down-converted photons which are simultaneously entangled in both polarization and transverse position-momentum.  Entanglement in position-momentum is very robust and it is, in fact,  actually difficult to produce down-converted photons pairs that are not entangled in position-momentum \footnote{C. H. Monken, private communication.}.  Most of the sources of polarization entangled photons, which generally rely on the superposition of two emission cones \cite{kwiat95,kwiat99}, already produce entanglement in position-momentum.    A difficulty arises in that this superposition is obtained in the far-field, so that by performing a position measurement by imaging one or both photons creates distinguishability in the form of ``which cone" information.    Thus, in order to utilize both position-momentum and polarization correlations in a QKD protocol, it is necessary to develop a position-momentum measurement scheme which does not destroy polarization entanglement.  Here we demonstrate experimentally that the quantum correlations
between position-momentum and polarization of photon pairs can be produced and
detected simultaneously. We then show that these simultaneous correlations can be used to implement QKD protocols based on 
four-dimensional states, which enhance the transmission rate per photon pair
and sensitivity to eavesdropping.  We provide three possible protocols: a parallel protocol with average transmission rate 1 bit/photon pair and error rate of $\frac{1}{4}$, a qu-quart protocol with average transmission rate 1 bit/photon pair and error rate of $\frac{3}{8}$, and a deterministic skewed qu-quart protocol with transmission rate 1 bit/photon pair and error rate $\frac{1}{4}$.      
\par 
 \begin{figure}
 \includegraphics[width=7cm]{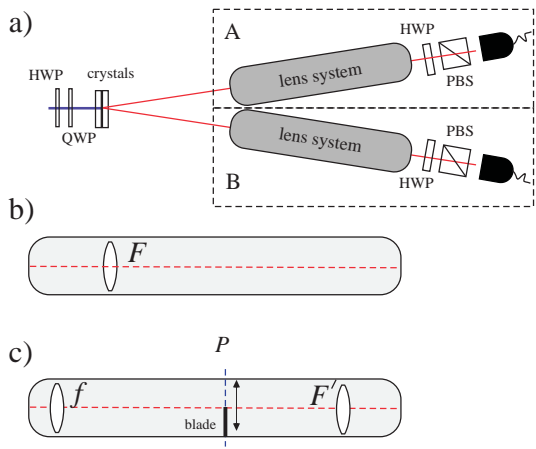}%
 \caption{\label{fig:1}a) Experimental setup for QKD. The PBS and HWP are used for polarization measurements.  b)  Lens system for momentum measurements.  See text.  c)  Lens system for position measurements.  See text. }
 \end{figure}
 The experimental setup is shown in Fig. \ref{fig:1}.  A He-Cd laser was used to pump two 2\,mm long Lithium Iodate crystals arranged with their optic axes perpendicular, creating down-converted signal and idler photons, as reported in Ref. \cite{kwiat99}.  The signal and idler photons were sent to detection systems $A$ and $B$ respectively.  Coincidence counts were registered using SPCM single photon counting modules equipped with $12$\,nm FWHM bandwidth interference filters and $0.3$\,mm diameter detection apertures.  Coincidence and single counts were registered using coincidence electronics and a personal computer. 
 \par
  It is well known that this source can be used to create photon pairs in a maximally entangled polarization state \cite{kwiat99}.  The polarization of the pump beam was adjusted using half- and quarter-wave plates so that the photon pairs were in the state $\ket{\phi^{-}}=(\ket{HH}-\ket{VV})/\sqrt{2}$.  With the lens systems removed, the polarization entanglement was initially verified by performing the usual polarization interference measurements using linear polarization analyzers consisting of half-wave plates (HWP) and polarizing beam splitters (PBS), where one of the analyzers is kept fixed at $45^\circ$ and the other is rotated.  Interference curves with visibilities above $95$ \% were typically observed in the coincidence counts, while the single count rates remained approximately constant, indicating a high degree of entanglement. 
\par
\begin{figure}
 \includegraphics[width=7cm]{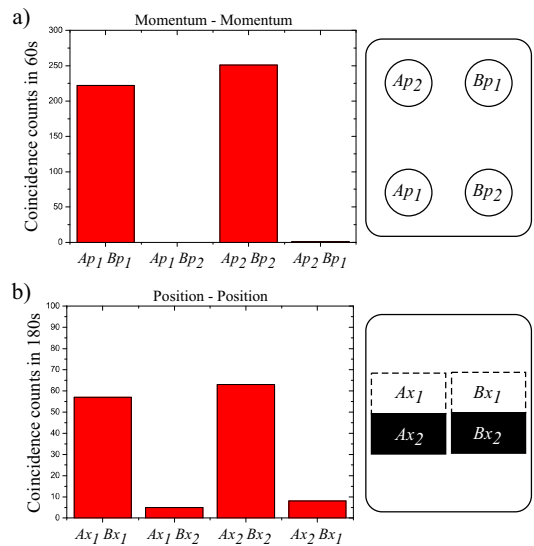}%
 \caption{\label{fig:res1}a) Experimental results showing correlations in a) momentum-momentum measurements and b) position-position measurements.  Also shown is a diagram of each measurement scheme (see text).}
 \end{figure}

Due to phase matching in SPDC, anti-correlation is observed in transverse momentum measurements.  That is, if one detects a signal photon with momentum $\brm{p}$, one finds an idler photon with momentum $-\brm{p}$.  One can measure momentum by creating the Fourier transform of the field at the source on the detection plane. This is achieved using a lens placed so that the image and object planes each lie in the focal plane of the lens, therefore mapping the transverse momentum distribution of the photons at the crystal face onto positions in the detection plane.  Fig. \ref{fig:1} b) shows our lens system consisting of a lens with focal length $F=500$\,mm.  The detectors were placed 1\,m from the source.  It is thus possible to choose detector positions that exhibit a high spatial anti-correlation.  Fig. \ref{fig:res1} a) shows coincidence counts for 2 sets of detector positions which display high momentum anti-correlations.  Positions $A_{p1}$ and $A_{p2}$ correspond to positions of detector $A$ and  likewise $B_{p1}$ and $B_{p2}$ for detector $B$.  Also shown is a diagram of the detector positions, from which it can be seen that when one photon is detected in the upper position, the other is detected in the lower position, illustrating the anti-correlation in momentum measurements.  The polarization analyzers were orientated at $\theta_{A}=-45^{\circ}$ and $\theta_{B}=45^{\circ}$ to ensure that we collected photons from both crystals.  
 \par
Position measurements can be performed by creating the image of the source at the detection plane \cite{howell04,dangelo04}.  Using an imaging system we see position correlations:  if the signal photon is detected at position $\brm{x}$, the idler photon is found at the same position $\brm{x}$.  Position correlations in down-converted photons are due to the fact that the photons are ``born" at the same position in the crystal.  However, simply imaging the crystal face will destroy the polarization entanglement present in this two-crystal source.  It has been shown that a two-crystal source displays interference that is analogous to a two-photon double-slit experiment \cite{ribeiro01}.  Thus imaging the crystal face provides  ``which-crystal" information that destroys polarization interference.  
 In order to overcome this difficulty, we developed a slightly more complex position measurement system, shown in Fig. \ref{fig:1} c).  A lens is used to image the crystal face at an intermediate plane $P$.  A thin blade is placed in plane $P$ such that it blocks half of the image, while a second lens placed in a Fourier configuration is then used to filter the spatial modes, which removes which path information.  Instead of moving the detector, the blade position was varied. For each detection system we defined positions $A_{x1}$, $A_{x2}$, $B_{x1}$ and $B_{x2}$ of the blade in the image plane $P$.  We emphasize that this two lens system does not perform the Fourier transform of the crystal face, since the blade removes half of the image. 
Using this detection system we measured the position correlations, which are shown in Fig. \ref{fig:res1} b).  The diagram shows the positions of the blades corresponding to each measurement, which shows that in this case we have position \emph{correlations} instead of anti-correlations as in the momentum-momentum case.
 \begin{figure}
 \includegraphics[width=7cm]{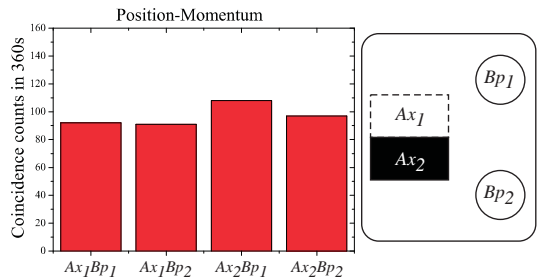}%
 \caption{\label{fig:res2} Experimental results for position-momentum measurements.  Also shown is a diagram of the measurement scheme (see text).}
 \end{figure}
\par
Since position and momentum are complementary observables, there is a complete absence of correlation when systems $A$ and $B$ perform measurements in different bases.  
Fig.  \ref{fig:res2} shows measurement results when $A$ implements a momentum measurement and $B$ implements a position measurement.  There is roughly the same detection probability for all detector combinations, which shows that there is no correlation between the detection positions.  
 \begin{figure}
 \includegraphics[width=7cm]{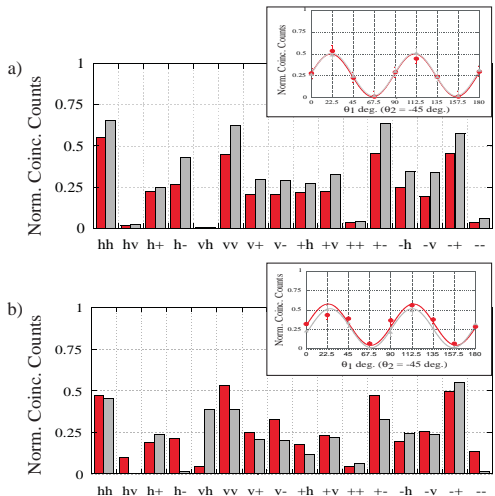}%
 \caption{\label{fig:res3} Polarization correlation measurements for  a) momentum measurements: $A_{p1}B_{p1}$ (red bars) and  $A_{p2}B_{p2}$ (grey bars) and b) position measurements $A_{x1}B_{x1}$ (red bars) and $A_{x2}B_{x2}$ (grey bars).  The insets show the polarization interference curves.}
 \end{figure}
 \begin{figure}
 \includegraphics[width=7cm]{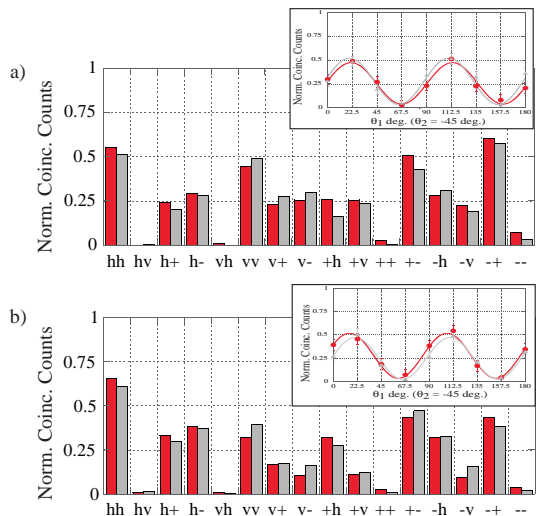}%
 \caption{\label{fig:res4}a) Polarization correlation measurements for  a) position-momentum measurements $A_{x1}B_{p1}$ (red bars) and $A_{x1}B_{p2}$ (grey bars), and b) position-momentum measurements $A_{x2}B_{p1}$ (red bars) and $A_{x2}B_{p2}$ (grey bars).  The insets show the polarization interference curves.}
 \end{figure}
 \par
In order to use polarization entanglement along with position-momentum correlations to establish a random keystring, it is necessary that the polarization state remains the same for all possible position-momentum measurement configurations.
Having established adequate position and momentum detection systems, we then performed polarization interference measurements in order to confirm that polarization correlations observed with different position-momentum measurement systems were the same.  Interference curves with visibilities ranging from 82\% to 92\% were observed.  Just as important as the interference visibility in this case is the relative phase of the polarization state, which remained approximately constant for all measurements.   
Fig. \ref{fig:res3} shows the polarization correlations when $A$ and $B$ both measure in the a) momentum and b) position basis.  Here we show coincidence counts for those detector combinations that result in nonzero coincidences. For combinations like $A_{p1}B_{p2}$ and $A_{x1}B_{x2}$, the number of coincidence counts were found to be much smaller than combinations like $A_{p1}B_{p1}$ and $A_{x1}B_{x1}$.
Fig. \ref{fig:res4} shows coincidence counts for those cases in which $A$ and $B$ measure in different bases. Fig. \ref{fig:res4} a) and b) show polarization correlations for the case where $A$ measures momentum while $B$ measures position.  The insets in FIG.'s \ref{fig:res3} and \ref{fig:res4} show polarization interference curves.  All experimental results show that polarization correlations can be observed in parallel to position-momentum correlations.  It is thus possible to use these DOF to implement QKD protocols.  
\par
Let us briefly analyze some possible protocols which could be implemented with position-momentum
and polarization. One first possibility is to run two parallel and independent BBM
protocols, one using polarization entanglement and other using position-momentum correlations\cite{almeida05}.  Since the two protocols are independent, Alice and Bob choose randomly between polarization bases as well as position and momentum bases.  In this case, an average transmission rate of one bit per photon pair is achieved, since $1/4$ of the time they choose the same basis in both DOF and establish 2 bits, $1/2$ of the time they choose the same basis in only one DOF and establish 1 bit, and $1/4$ of the time they choose different bases in both DOF.  
The error rate due to intercept-resend eavesdropping is the same as a single BBM protocol: $\frac{1}{4}$. 
\par
A second possibility  is encoding in 
qu-quarts instead of qubits. A four-dimensional alphabet can be defined using position-momentum and polarization. For instance, let us define $a \equiv H,X_1$, which stands for one photon with linear polarization $H$ at position $X_1$, $b \equiv H,X_2$, $c \equiv V,X_1$ and $d \equiv V,X_2$.  A complementary alphabet would be: $\alpha \equiv +,P_1$, which stands for one photon with linear polarization $+45^{\circ}$ and momentum $P_1$, $\beta \equiv +,P_2$, $\gamma \equiv -,P_1$ and $\delta \equiv -,P_2$.  Alice and Bob then both choose between two sets of measurements:  the roman letter basis, consisting of $H/V$ polarization and $X$ measurements, or the greek letter basis, comprised of $+/-$ polarization and $P$ measurements.  Here the transmission rate is again 1 bit per photon pair, since $1/2$ of the time they choose the same basis and establish 2 random bits, while the other $1/2$ of the time they choose different bases and establish no random bits.  
In addition to the increase in the transmission rate, the error rate due to
intercept-resend eavesdropping would be improved from $\frac{1}{4}$ to $\frac{3}{8}$, since Eve chooses the wrong basis with probability $1/2$ and consequently has a $3/4$ probability to send the wrong state.   
\par A third possibility is for Alice and Bob to skew their combined measurements.  For example, suppose Alice chooses between (\text{i}) $H/V$ polarization and $X$ measurements and (\textit{ii}) $+/-$ polarization and $P$ measurements, while Bob chooses between (\text{i}) $H/V$ polarization and $P$ measurements and (\textit{ii}) $+/-$ polarization and $X$ measurements.  Skewing the measurement bases guarantees that Alice and Bob always agree on a basis in one DOF.  This protocol is deterministic, in the sense that Alice and Bob establish one random bit for every photon pair sent, and thus every photon pair contributes to the sifted key.  
\par  
We note that this last example is similar to a recent proposal for deterministic QKD where security is provided by AVN tests of local realism \cite{pan05}.   In the protocol proposed in Ref. \cite{pan05}, Alice and Bob choose randomly between three groups of measurements. Two of the measurement groups are analogous to the skewed basis ab ove.  It is possible
to implement QKD based on AVN with polarization and position-momentum, however, our experimental results are not sufficient to demonstrate such a protocol.   One measurement required is a
single-photon Bell state measurement, which could be realized with linear
optics \cite{pan05}. The transmission rate QKD based on AVN is still one bit
per photon pair. The error rate was not analyzed in Ref.\cite{pan05}, but it
we conjecture that it would be increased, since the three groups of measurements are
analogous to three basis.
\par
In summary, we have experimentally observed the simultaneous entanglement
between polarization and position-momentum of photon pairs from 
parametric down-conversion. Our results show that four-dimensional quantum key
distribution can be implemented using these degrees of freedom, enhancing
the transmission rate and error rate induced by eavesdropping.

 \begin{acknowledgments}
The authors would like to acknowledge fruitful discussions with  L. Davidovich,  K. Dechoun, R. L. de Matos Filho, D. Jonathan, A. Z. Khoury and C. H. Monken. Financial support was provided by the Brazilian funding agencies CNPq, CAPES and FAPERJ and the Brazilian Millennium Institute for Quantum Information.
\end{acknowledgments}

\end{document}